\documentclass{article}

\usepackage[final,nonatbib]{neurips_2023}

\usepackage[utf8]{inputenc} 
\usepackage[T1]{fontenc}    
\usepackage{hyperref}       
\usepackage{url}            
\usepackage{booktabs}       
\usepackage{amsfonts}       
\usepackage{nicefrac}       
\usepackage{microtype}      
\usepackage{xcolor}         

\usepackage{wrapfig}
\usepackage{times}  
\usepackage{helvet}  
\usepackage{courier}  
\usepackage{graphicx} 
\usepackage{caption} 
\usepackage{xparse}
\usepackage{graphics}
\usepackage{subcaption}
\usepackage{makecell}
\usepackage{epstopdf}
\usepackage{rotating,multirow}
\usepackage{amsmath,amssymb} 
\usepackage{color}
\usepackage{wrapfig,lipsum}
\usepackage{tabularx}
\usepackage{algorithm}
\usepackage{algorithmic}
\usepackage[algo2e]{algorithm2e}
\usepackage{mathtools}
\usepackage{cite}

\newcommand{\etal}{\textit{et al. }}

\newcommand{\ie}{\textit{i.e., }}
\newcommand{\aka}{\textit{a.k.a. }}

\newcommand{\defeq}{\coloneqq}

\newcommand{\E}{\mathbb{E}}

\newcommand{\Ea}[1]{\E\left[#1\right]}
\newcommand{\Eb}[2]{\E_{#1}\!\left[#2\right]}

\newcommand{\bI}{\mathbf{I}}

\newcommand{\bzero}{\mathbf{0}}

\newcommand{\bx}{\mathbf{x}}

\newcommand{\bepsilon}{{\boldsymbol{\epsilon}}}
\newcommand{\bmu}{{\boldsymbol{\mu}}}

\newcommand{\bSigma}{{\boldsymbol{\Sigma}}}

\title{Adversarial Denoising Diffusion Model for Unsupervised Anomaly Detection}

\author{%
  Jongmin Yu$^{1}$, Hyeontaek Oh$^{2}$, Jinhong Yang$^{3}$\\
  $^{1}$Department of Applied Mathematics and Theoretical Physics, University of Cambridge, \\ Centre for Mathematical Sciences, Wilberforce Rd, Cambridge CB3 0WA, United Kingdom \\
  $^{2}$Institute for IT Convergence, Korea Advanced Institute of Science and Technology (KAIST), \\Daejeon, 34141, Rep. of Korea \\
  $^{3}$Department of Medical Information Technology, Inje University, \\ Kimhae, Republic of Korea, 50834\thanks{Email to the corresponding author: jinhong@inje.ac.kr}  \\
  \texttt{jy522@cam.ac.uk, hyeontaek@kaist.ac.kr, jinhong@inje.ac.kr}
}

\begin{document}

\maketitle

\begin{abstract}
In this paper, we propose the Adversarial Denoising Diffusion Model (ADDM). The ADDM is based on the Denoising Diffusion Probabilistic Model (DDPM) but complementarily trained by adversarial learning. The proposed adversarial learning is achieved by classifying model-based denoised samples and samples to which random Gaussian noise is added to a specific sampling step. With the addition of explicit adversarial learning on data samples, ADDM can learn the semantic characteristics of the data more robustly during training, which achieves a similar data sampling performance with much fewer sampling steps than DDPM. We apply ADDM to anomaly detection in unsupervised MRI images. Experimental results show that the proposed ADDM outperformed existing generative model-based unsupervised anomaly detection methods. In particular, compared to other DDPM-based anomaly detection methods, the proposed ADDM shows better performance with the same number of sampling steps and similar performance with 50\% fewer sampling steps.
\end{abstract}

\section{Introduction}
\label{sec:1}
The diffusion model is one of the generative models based on thermal non-equilibrium physics \cite{pmlr-v37-sohl-dickstein15}. Compared with other generative models such as Generative Adversarial Network (GAN) \cite{NIPS2014_5ca3e9b1} and Variational Autoencoder (VAE) \cite{kingma2022autoencoding}, the diffusion model generates more high-quality data samples, \ie the diffusion models can derive more discriminative parametric distribution for a given dataset. Based on their superior capability in sampling data, the diffusion models are establishing remarkable success in various domains, such as image and audio generation \cite{NEURIPS2022_ec795aea,Deng_2023_CVPR,Kawar_2023_CVPR} and natural language processing \cite{pmlr-v202-lin23d}. Those achievements triggered various unsupervised anomaly detection (AD) studies that apply diffusion models to detect data anomalies \cite{Wyatt_2022_CVPR,code_paper2,code_paper3,Graham_2023_CVPR}.

Those unsupervised AD studies using diffusion models \cite{Wyatt_2022_CVPR,code_paper2,code_paper3,Graham_2023_CVPR} have justified that the diffusion models can be used for unsupervised AD and can achieve competitive performance compared with other generative model-based AD methods. The high-quality data sampling capability of the diffusion models can be a strong advantage in driving unsupervised AD methods. However, applying diffusion models is still challenging due to high computational costs caused by iterative sampling or the unstable quality of generated data depending on the number of samples \cite{croitoru2023diffusion}.

This paper introduces the Adversarial Denoising Diffusion Model (ADDM), an end-to-end unsupervised AD method using a diffusion model. The proposed ADDM minimises the noise prediction error and explicitly minimises adversarial loss about the denoised sample. Compared to other diffusion model-based methods, ADDM can dramatically reduce the number of data sampling (reverse process) steps by providing an explicit process to improve the quality of denoised data. The ADDM outperforms other diffusion model-based AD methods \cite{code_paper3,Wyatt_2022_CVPR} with 6.2\% better performance in our experiments using MRI images for brain tumour detection.

\section{Preliminaries}
The goal of diffusion models \cite{pmlr-v37-sohl-dickstein15} is to find the parameterised data distribution $p_\theta(\bx_0)$ using a given data $\bx_0 \sim q(\bx_0)$. To do this, when a data $\bx_0$ is given, the training of diffusion models conducts a \emph{forward process} (\aka diffusion process) $q(\bx_{t} | \bx_{t-1})$, which adds Gaussian noise to the data and a \emph{reverse process} (\aka denoising process) $p_\theta(\bx_{t-1}|\bx_t)$, which denoises the given noised data by subtracting predicted noise.

The forward process $q(\bx_{t} | \bx_{t-1})$ is a task to add Gaussian noise to data at a certain time step $t \leq T$. The Gaussian noise is generated by a Gaussian probability $\mathcal{N}(\cdot)$ with scheduled variance: $\beta_1, \dotsc, \beta_T$. The entire forward process to generate completely noised sample $\bx_{T}$ is represented by 
\begin{equation}
\begin{aligned}
q(\bx_{1:T} | \bx_0) &\defeq \prod_{t=1}^T q(\bx_t | \bx_{t-1} ), \qquad 
q(\bx_t|\bx_{t-1}) \defeq \mathcal{N}(\bx_t;\sqrt{1-\beta_t}\bx_{t-1},\beta_t \bI).
\end{aligned}
\label{eq:forwardprocess}
\end{equation}

The reverse process $p_\theta(\bx_{t-1}|\bx_t)$ can be considered as a denoising task. For each time step $t$, a diffusion model predicts a noise and subtracts it from the noised data. This task is represented by Markov Chain so that it can be represented by parametric conditional distribution, as follows:
\begin{equation}
\begin{aligned}
  p_\theta(\bx_{0:T}) &\defeq p(\bx_T)\prod_{t=1}^T p_\theta(\bx_{t-1}|\bx_t), \qquad 
  p_\theta(\bx_{t-1}|\bx_t) \defeq \mathcal{N}(\bx_{t-1}; \bmu_\theta(\bx_t, t), \bSigma_\theta(\bx_t, t)).
\end{aligned}\
\label{eq:reverseprocess}
\end{equation}

The learning of the diffusion model is to find the suitable parametric distribution $p_\theta(\bx_{0})$ representing a given data. However, the negative log-likelihood of $p_\theta(\bx_{0})$ is not nicely computable, so it is alternatively optimised by variational lower-bound, which is represented by
\begin{equation}
\begin{aligned}
L_{dm}  \defeq  \Ea{-\log p_\theta(\bx_0)} \leq \Eb{q}{ - \log \frac{p_\theta(\bx_{0:T})}{q(\bx_{1:T} | \bx_0)}}
  = \mathbb{E}_q\bigg[ -\log p(\bx_T) - \sum_{t \geq 1} \log \frac{p_\theta(\bx_{t-1} | \bx_t)}{q(\bx_t|\bx_{t-1})} \bigg].
\end{aligned}
\label{eq:dm_loss}
\end{equation}

In DDPM \cite{ho2020denoising}, it found that the forward and reverse processes are represented by reparameterisation tricks; therefore, it can replaced by the minimisation of prediction error for a Gaussian noise and noise prediction obtained by a neural network. As a result, Eq. \eqref{eq:dm_loss} is more simplified to the $l2$-distance between the Generated Gaussian noise $\bepsilon \sim \mathcal{N}(\bzero, \bI)$ and predicted noise $\bepsilon_\theta$ at a certain time step. Using the notations, $\alpha_t \defeq 1-\beta_t$ and $\bar\alpha_t \defeq \prod_{s=1}^t \alpha_s$, the simplified loss is represented by 
\begin{equation}
\begin{aligned}
    \Eb{\bx_0, \bepsilon}{ \frac{\beta_t^2}{2\sigma_t^2 \alpha_t (1-\bar\alpha_t)}  \left\| \bepsilon - \bepsilon_\theta(\sqrt{\bar\alpha_t} \bx_0 + \sqrt{1-\bar\alpha_t}\bepsilon, t) \right\|^2}.
\end{aligned}
\label{eq:ddpm_loss}
\end{equation}

\section{Adversarial Denoising Diffusion Model}
Unlike GAN \cite{NIPS2014_5ca3e9b1} or VAE \cite{kingma2022autoencoding}, which directly generate data from relatively low-dimensional Gaussian noise and classify whether the samples are generated or given (in the case of GAN) or minimise Euclidean distances of generated samples and a given sample (in the case of VAE), to generate data from complete, from a given noise which having the same dimensionality with the data, diffusion models repeatedly performed the reverse process, \ie predicting the noise present in the given data and remove it. 

The diffusion model methodology, that is, generating data through an iterative procedure that gradually restores data from the same dimensional noise, shows outstanding data sampling performance, but compared to GAN and VAE, the sampling process of the diffusion model is cost-efficient. Additionally, the quality of generated data is varied depending on the number of sampling steps \cite{croitoru2023diffusion}. This issue is because diffusion model learning involves noise prediction, which is not closely related to the semantic properties of data. When generating noise, $\bx_0$ is used conditionally (See Eq. \eqref{eq:reverseprocess}), but the model ultimately learns the minimum prediction error between Gaussian noise and the noise present in the image (See Eq. \eqref{eq:ddpm_loss}). In this study, adversarial learning is proposed to minimise changes in the diffusion model's structure and preserve the semantic characteristics of data.

\begin{figure}[t]
 \includegraphics[width=\textwidth]{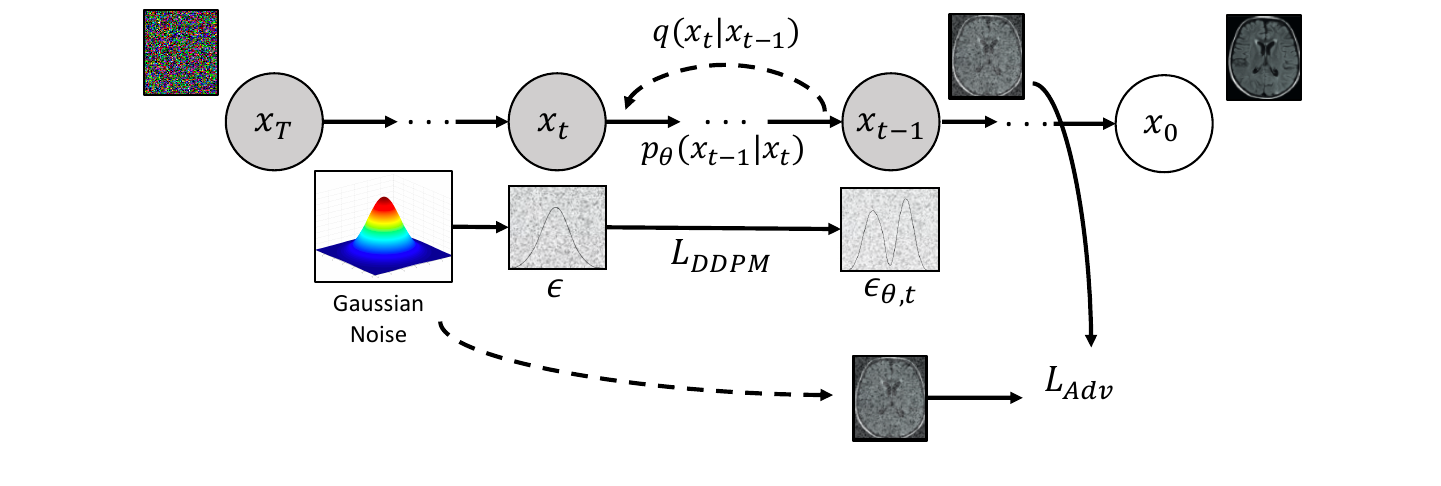}
\caption{\small Illustration for Learning of Adversarial Denoising Diffusion Model (ADDM). $L_\mathrm{DDPM}$ denotes the loss function to minimise the $l2$-distance between the generated error from Gaussian distribution and the predicted error by a model. $L_\mathrm{Adv}$ indicates the adversarial loss between the model-based denoised sample $x_{t-1}$ and virtually generated noised samples, for the certain time step $t-1$.}
  \label{fig:archs}
\vspace{-2ex}
\end{figure}

Figure \ref{fig:archs} shows the architectural details of the proposed ADDM. The objective function of the adversarial denoising diffusion model (ADDM) is defined by the summation of the DDPM loss (Same with Eq. \eqref{eq:ddpm_loss}) and the additional adversarial loss using a balancing weight $\lambda$, represented by
\begin{equation}
\begin{aligned}
  L_\mathrm{ADDM} & \defeq \underbrace{\Eb{t, \bx_0, \bepsilon}{ \left\| \bepsilon - \bepsilon_\theta(\sqrt{\bar\alpha_t} \bx_0 + \sqrt{1-\bar\alpha_t}\bepsilon, t) \right\|^2}}_{L_\mathrm{DDPM}} 
  \\& +\underbrace{\lambda \Eb{t, \bx_0, \bepsilon}{\log{}D(q(\bx_{t-1}))} + \Eb{t, \bx_0, \bepsilon}{\log{}(1-D(p_\theta(\bx_{t-1}|\bx_t))}}_{L_\mathrm{Adv}},
\end{aligned}
\label{eq:addm_loss}
\end{equation}
where $D$ denotes the discriminator to apply the adversarial learning on ADDM. $q(\bx_t|\bx_{t-1})$ defines the $t$-step noised data obtained by the forward process (Eq. \eqref{eq:forwardprocess}), and $p_\theta(\bx_{t-1}|\bx_t)$ indicates the denoised data produced by the reverse process (Eq. \eqref{eq:reverseprocess}).

$L_\mathrm{DDPM}$ is a simplified version of Eq. \eqref{eq:ddpm_loss}. DDPM \cite{ho2020denoising} supposes that the simplified loss function is more beneficial to the quality of the sample and implementation efficiency; therefore, we keep this loss format in our study. $L_\mathrm{Adv}$ is the loss term for the adversarial learning about the denoised data. $L_\mathrm{Adv}$ tries to distinguish between the real noise data $q(\bx_t|\bx_{t-1})$ and the denoised data $p_\theta(\bx_{t-1}|\bx_t)$. 

In predicting a noise for the noise removal process for data generation through DDPM, the ADDM model can learn semantic features of the data and information about the noise to be predicted. Through this direct learning of semantic features of data, we expect to reduce the number of sampling steps required to generate sufficient quality data. Besides, the adversarial loss helps to reduce blurriness, which means it helps to learn high-frequency features to add more local details in denoising the data. This property will help prove the accuracy of AD methods by reducing the false-positive results. Through ablation studies, we will demonstrate that the proposed adversarial learning helps AD.

\section{Experiment}
\subsection{Experimental settings}
\textbf{Dataset and Protocol:}
We utilise the two datasets. The first dataset is the Neurofeedback Skull-Stripped (NFBS) repository \cite{puccio2016preprocessed}. The repository provides 125 MRI images captured from normal people, so there are no anomalies in the MRI images. The second dataset contains 22 T1-weighted MRI scans provided by the Centre for Clinical Brain Sciences from the University of Edinburgh \cite{pernet2016structural}. The second dataset provides MRI images containing brain tumours. To follow an experimental protocol for the unsupervised AD, the training dataset has to be composed of normal samples only. We train the ADDM using the first dataset and conduct AD experiments using the second dataset. We refer to the experiment protocol of AnoDDPM \cite{Wyatt_2022_CVPR}.

\noindent
\textbf{Implementations:} The resolution of images is resized to 256 $\times$ 256. Adam optimiser \cite{kingma2014adam} is used for the optimisation algorithm. The balancing weight $\lambda$ is set by 0.05. The number of epochs and the batch size are set by 3000 and 4, respectively. To demonstrate the effectiveness of adversarial learning, we evaluate the AD performance with 300, 500, and 1000 sampling steps ($T$). The learning rate is initialised by 0.0001, and it is decayed by multiplying 0.999 for every 200 epoch.

\subsection{Experimental results}
\noindent
\textbf{Effectivenss of $L_\mathrm{Adv}$:} We train the ADDM with 300, 500, and 1000 sampling steps. Table \ref{tbl:results} contains the quantitative performances of the ADDM depending on $T$. The ADDM obtains the best performance with 1000 sampling steps. It produces 0.403 Dice and 0.917 AUC. The overall performance of the ADDM with 1000 sampling steps is better than the other two ADDMs trained with 300 and 500 sampling steps. However, the ADDMs trained with 300 and 500 sampling steps produce competitive performance compared with other DDPM-based models \cite{code_paper3,Wyatt_2022_CVPR}. In particular, the ADDM trained with 500 sampling steps outperforms the Pinaya \etal \cite{code_paper3}, which is structurally equivalent to the DDPM with 1000 sampling steps. The experimental results justify that the adversarial learning on the ADDM improves the robustness of the diffusion models with respect to the sampling step.

\noindent
\textbf{Comparison with SOTA methods:} We compare the proposed method with various generative model-based AD methods \cite{baur2021autoencoders,schlegl2019f,pinaya2021unsupervised,Wyatt_2022_CVPR}. Table \ref{tbl:results} shows the quantitative results on the dataset. Listed methods have been chosen for performance comparison:  Autoencoder (AE), Variational AE,\cite{baur2021autoencoders}, f-AnoGAN \cite{schlegl2019f}, Transformer \cite{pinaya2021unsupervised}, Pinaya \etal \cite{code_paper3}, and AnoDDPM \cite{Wyatt_2022_CVPR}. In particular, Pinaya \etal \cite{code_paper3} and AnoDDPM \cite{Wyatt_2022_CVPR} are built based on DDPM \cite{ho2020denoising}, so their baseline methods are similar to the ADDM. Both approaches compile reconstruction-based AD methods for MRI images using diffusion models. Interestingly, the architectural details of Pinaya \etal \cite{code_paper3} are almost identical to the DDPM \cite{ho2020denoising}. AnoDDPM \cite{Wyatt_2022_CVPR} is built based on the DDPM and uses a new noising approach called \emph{Simplex Noise}.

The quantitative results in Table \ref{tbl:results} show that the ADDM outperforms other methods. The ADDM ($T$=1000) produces 0.917 AUC, which is 6.2\% higher performance than the second-ranked method (AnoDDPM). Moreover, the proposed ADDM trained with 500 sampling steps also achieves competitive performance with the AnoDDPM that requires 1000 sampling steps. This result shows that the proposed ADDM is more cost-efficient than AnoDDPM. 
Overall, the experimental results show that the proposed adversarial loss improves AD performances and outperforms existing SOTA AD detection methods on MRI images.

\begin{figure}[t]
\centering
\begin{minipage}[t]{0.8\linewidth}
\resizebox{\linewidth}{!}{%
\begin{tabular}{lccccc}
    \hline
      \bfseries Method & Dice  & AUC  & IoU  & Precision  & Recall \\
      \hline
      \hline
      AE (Spatial) \cite{baur2021autoencoders}                   & 0.252     & 0.707  & 0.162 & 0.258 & 0.279 \\
      VAE (Dense) \cite{baur2021autoencoders}                    & 0.317     & 0.734  & 0.203 & 0.297 & 0.313\\
      f-AnoGAN \cite{schlegl2019f}                               & 0.128     & 0.789  & 0.093 & 0.362 & 0.080 \\
      Transformer \cite{pinaya2021unsupervised}                  & 0.241     & 0.695  & 0.193 & 0.275 & 0.120 \\
      Pinaya \etal \cite{code_paper3} ($T=1000$)                                 & 0.375     & 0.815  & 0.238 & 0.367 & 0.452\\
      AnoDDPM \cite{Wyatt_2022_CVPR} ($T=1000$)                                  & 0.383     & 0.863  & 0.269 & 0.373 & 0.468 \\
        \hline
      ADDM ($T=300$)                                & 0.301     & 0.811  & 0.213 & 0.293 & 0.369  \\
      ADDM ($T=500$)                               & 0.379     & 0.861  & 0.271 & 0.361 & 0.491 \\
      ADDM ($T=1000$)                              & \textbf{0.403}     & \textbf{0.917}  & \textbf{0.289} & \textbf{0.392} & \textbf{0.508} \\
      \hline
    \end{tabular}
    }
    \captionof{table}{\small Quantitative performance comparison of unsupervised AD methods based on generative models. $T$ denotes the number of sampling steps of a diffusion model. 
    }
        \label{tbl:results}
\end{minipage}
\vspace{-2ex}
\end{figure}

\section{Conclusion}
In this work, we have presented an adversarial denoising diffusion model (ADDM) for AD on MRI images, which can derive a more discriminative AD method based on diffusion models. Similar to DDPM, the proposed ADDM learns a noise prediction model based on the forward and reverse processes. The adversarial learning is applied to improve the quality of denoised data explicitly. Our ablation studies have shown that adversarial learning is helpful not only in improving AD performance but also in reducing the sampling step to obtain high-quality data. In comparison with existing AD methods based on generative models, ADDM outperforms existing SOTA AD methods.

\section{Acknolwdgement}
This work was partly supported by the Institute of Information \& Communications Technology Planning \& Evaluation(IITP) grant funded by the Korea government(Ministry of Science and ICT; MSIT) (2020-0-00048, Development of 5G-IoT Trustworthy AI-Data Commons Framework) and by the National Research Foundation of Korea (NRF) grant funded by the Korea government (MSIT) (NRF-2022R1C1C2003437)

\bibliographystyle{unsrt}
{\small \bibliography{mybib}}
\end{document}


\bibliographystyle{abbrvnat}

\maketitle

\appendix

\vspace{-10mm}

Here, we present additional experiments and analyses that could not be included in the main text due to space constraints. All figures and references in this supplementary file are self-contained.
The contents included in these supplementary materials are as follows: 
1) The description of base anomaly detectors, 2) IADC Algorithm, 3) The effect of resampling, 4) Anomaly detection on unseen data, 5) Sensitivity analysis of the outlier ratio $r$, 6) Variance analysis, and 7) Feature distribution visualization.



\section{Base anomaly detectors}





In this section, we list details of each base anomaly detector used in our experiments.
Let a given base anomaly detection model $\calD$ be trained with the objective function $\calL_{\calD}(X)$ on the training dataset $X=\{x_1, \cdots, x_n\}$. Then, the $j^{th}$ base model $\calD_j, j\in\{1,\cdots,m\}$ of the ensemble $\calE$ is iteratively trained with the objective function $\calL_{\calD_j}(X, W^t)$, where $t$ is an index of the iterative learning stage and $W^t=\{w_1^t, \cdots, w_n^t\}$ is a set of importance weights of the training dataset at iteration $t$, where $w^0_i=1$. 
%
%
%
In the following, details of objective function $\calL_{\calD}(X)$, objective function for IADC $\calL_{\calD_j}(X,W^t)$, and anomaly score $s$ are described for each base anomaly detection model: OC-SVM~\cite{scholkopf2001estimating}, AE~\cite{sakurada2014anomaly}, hybrid OC-SVM~\cite{andrews2016detecting}, Deep SVDD~\cite{ruff2018deep}, RSRAE~\cite{lai2020robust}, and DAGPR~\cite{fan2020robust}.

\paragraph{OC-SVM}
The objective of OC-SVM finds a maximum margin hyperplane in feature space $\calF$ that best separates the data from the origin.  
Given a dataset $X, x_i \in \calX$, OC-SVM solves the following quadratic problem. Here, $\calW$, $\rho$ denote a weight vector, and an offset parameterizing a hyperplane in $\calF$ associated with the kernel. $\Psi$ is a feature map $\calX \rightarrow \calF$ such that the inner product in the image of $\Psi$ can be computed by evaluating some single kernel. $\xi$ is a nonnegative slack variable allowing the margin to be soft and $\nu$ controls the trade-off in the objective.
$$
\begin{aligned}
&\min_{\calW\in F, \xi\in\mathbb{R}^n, \rho\in\mathbb{R}} \frac{1}{2} \norm{\calW}^2 + \frac{1}{\nu n} \sum_{i}^{n} \xi_i - \rho \\
&\textrm{subject to} \; (\calW\cdot \Psi(x_i)) \geq \rho - \xi_i, \; \xi_i \geq 0\\
\end{aligned}
$$

For consistency with deep learning-based methods, we represent $\calL_{\calD}(X)$ and $\calL_{\calD_j}(X,W^t)$ of OC-SVM as follows.
\begin{equation}  
    \label{eqn:ocsvm}
    \begin{aligned}
    \calL_{\calD}(X) = &\frac{1}{2} \norm{\calW}^2 + \frac{1}{\nu n} \sum_{i}^{n} \xi_i - \rho \\
    &\textrm{subject to} \; (\calW\cdot \Psi(x_i)) \geq \rho - \xi_i, \; \xi_i \geq 0\\
\end{aligned}
\end{equation}
 
\begin{equation}  
    \label{eqn:ocsvm-iadc}
    \begin{aligned}
    \calL_{\calD_j}(X, W^t) = &\frac{1}{2} \norm{\calW_j}^2 + \frac{1}{\nu n} \sum_{i}^{n} w_i^t \: \xi_{i,j} - \rho_j \\
    &\textrm{subject to} \; (\calW_j\cdot \Psi(x_i)) \geq \rho_j - w_i^t \: \xi_{i,j}, \; \xi_{i,j} \geq 0\\
\end{aligned}
\end{equation}

$$s_{i,j} = -(\calW_j \cdot \Psi(x_i) - \rho_j)$$


\paragraph{Autoencoder}
The objective function of AE is reconstruction loss between the input and output of the model. 
AE is applied to anomaly detection under the assumption that data has variables correlated with each other and can be embedded into a lower dimensional subspace, in which normal and anomalous samples appear significantly different. The reconstruction error term represents the anomaly score; thereby, the reconstruction loss is directly used as an anomaly objective. 
Here, $M$ and $G$ denote an encoder and a decoder parameterized by $\Theta$ and $\Phi$, respectively.

%
\begin{equation}
    \label{eqn:AE}
    \calL_{\calD}(X) ={\frac{1}{n}} \sum_{i=1}^{n} \norm{x_i - G(M(x_i; \Theta); \Phi)}^2
\end{equation}

\begin{equation}
    \label{eqn:AE-iadc}
    \calL_{\calD_j}(X, w^t) ={\frac{1}{n}} \sum_{i=1}^{n} w_i^t \: \norm{x_i - G_j(M_j(x_i; \Theta_j); \Phi_j)}^2
\end{equation}

$$s_{i,j} = \norm{x_i - G_j(M_j(x_i; \Theta_j); \Phi_j)}^2$$


\paragraph{hybrid OC-SVM}
In hybrid OC-SVM, OC-SVM is trained with latent features of a dataset extracted by an AE. When hybrid OC-SVM is applied to IADC, both AE and OC-SVM are iteratively trained with importance weights as described in \Eref{eqn:AE-iadc} and \Eref{eqn:ocsvm-iadc}. Then, the importance weights are updated by the anomaly score of OC-SVM.

\paragraph{Deep SVDD}
Inspired by kernel-based one-class classification and minimum volume estimation, Deep SVDD trains a neural network while minimizing the volume of a hypersphere that encloses the network representations of the dataset. 
The objective function of Deep SVDD is as follows, where $\phi$ denotes a neural network with parameters $\mathcal{W}$ and maps input data $x_i$ to a feature space $\calF$. $c$ denotes a center of a hypersphere in $\calF$.
Here, pretraining with AE is applied to initialize the parameter $\mathcal{W}$, but importance weight is not applied during pretraining.
%
\begin{equation}  
    \label{eqn:DSVDD}
    \calL_{\calD}(X) = {\frac{1}{n}} \sum_{i=1}^{n} \norm{\phi(x_i; \mathcal{W}) - c}^2 
\end{equation}

\begin{equation}
    \label{eqn:DSVDD-iadc}
    \calL_{\calD_j}(X, w^t) = {\frac{1}{n}} \sum_{i=1}^{n} w_i^t \: \norm{\phi_j(x_i; \mathcal{W}_j) - c_j}^2 
\end{equation}

$$s_{i,j} = \norm{\phi_j(x_i; \mathcal{W}_j) - c_j}^2.$$



\paragraph{RSRAE} 
RSRAE is a neural network for unsupervised anomaly detection with a robust subspace recovery layer (RSR layer) combined within an AE.
In RSRAE, the encoder maps each data point $x_i$ to its latent code $z_i$ with dimension $D$. Then, $z_i$ is linearly transformed to $\Tilde{z_i}$ with a lower dimension $d$ by the RSR layer. The decoder maps $\Tilde{z_i}$ to $\Tilde{x_i}$ in the original space. 
These forward maps are represented as follows. 
%
%
$$z_i = M(x_i; \Theta), \quad \Tilde{z_i} = A z_i, \quad \Tilde{x_i} = G(\Tilde{z_i}; \Phi)$$

where $M$ is an encoder with parameters $\Theta$, $A$ is a linear projection matrix, and $G$ is a decoder with parameter $\Phi$. 
The RSR layer, $A \in \mathbb{R}^{d\times D}$, seeks to extract the underlying subspace from a latent representation and removes outliers that lie away from this subspace. The objective function of RSRAE consists of reconstruction loss from the autoencoder and the loss from the RSR layer. The second and third terms in~\Eref{eqn:RSRAE} are motivated by the original RSR problem~\cite{lerman2018overview}. We apply the importance weights on the data reconstruction term and the projection reconstruction term (\Eref{eqn:RSRAE-iadc}).


\begin{equation}  
    \label{eqn:RSRAE}
    \calL_{\calD}(X) = \sum_{i=1}^{n} \norm{x_i - \Tilde{x_i}}_2 
                    + (\lambda_1 \sum_{i=1}^{n} \norm{z_i-A^TA z_i}_2
                    + \lambda_2 \norm{AA^T - I_d}_F^2)
\end{equation}

\begin{equation}   
    \label{eqn:RSRAE-iadc}
    \calL_{\calD_j}(X, w^t) = \sum_{i=1}^{n} w_i^t \: 
                            (\norm{x_i - \Tilde{x_i}}_2 + \lambda_1 \norm{z_i-A^TA z_i}_2 )
                            + \lambda_2 \norm{AA^T - I_d}_F^2
\end{equation}

$$s_{i,j} = \norm{x_i - G_j(A_j(M_j(x_i;\Theta_j));\Phi_j)}^2$$


\paragraph{DAGPR}
DAGPR is a hybrid unsupervised anomaly detection method, which integrates deep convolutional autoencoder (DCAE) and Gaussian process regression (GPR). 
In DAGPR, DCAE is used to extract features and reduce GPR computation by compressing the data. The objective function of DCAE can be represented as \Eref{eqn:AE}. 
Then the latent features of the encoder and the corresponding reconstruction error terms are combined and fed into GPR with pseudo-label assignments. 
GPR is iteratively trained, refining the predictions of the current stage based on the previous stage predictions. 
The sample distribution mean of GRP is used as an anomaly score. For the initial pseudo-label assignment, the reconstruction error is used. For more details of the process, refer to the work~\cite{fan2020robust}.
When we apply IADC to DAGPR, the importance weights are only applied to DCAE as described in \Eref{eqn:AE-iadc}.


\section{IADC learning process}

We present the detailed learning process of our framework IADC. The anomaly detectors can be categorized into two groups; ones that require AE pretraining, \eg OC-SVM hybrid, Deep SVDD, DAGPR, and the ones that do not require AE pretraining, \eg OC-SVM, AE, RSRAE. The former group of approaches optimizes objectives that are not related to a reconstruction task, while the latter group of approaches optimizes objectives related to a reconstruction task, \eg AE, RSRAE, or does not use deep representations, \eg OC-SVM.

\begin{algorithm}[h!]
	\caption{Training anomaly detectors with IADC}
	\label{alg:train}
	\textbf{Input:} Dataset $X$, Model weights $\calW_j$, Encoder weights $\Theta_j$ and Decoder weights $\Phi_j$, $ j\in\{1, \cdots, m\}$\\
    \textbf{Input:} Hyperparameters: $m$, $\text{epochs}_{AE}$, $\text{epochs}_{\calD}$, $\alpha_{AE}$ and $\alpha_{\calD}$
	\begin{algorithmic}[1]
        \Initialize{\strut$w_{i} \gets 1$, $i=1,\ldots,n$}
        \Repeat
            \State Data resampling $X_1,\cdots,X_m$
            \If{Require pretrain}
                \For {$j=1,\ldots,m$}
                    \For {$e=1,\ldots, \text{epochs}_{AE}$}
                        \State Compute $\calL_{AE}$ by \Eref{eqn:AE-iadc}
                        \State $\Theta_j \gets \Theta_j - \alpha_{AE} \nabla\calL_{AE}$
                        \State $\Phi_j \gets \Phi_j - \alpha_{AE} \nabla\calL_{AE}$
                    \EndFor
                \EndFor
                \State Initialize $\calW$ with $\Theta$ or use the encoder as a feature extractor
            \EndIf
            \For {$j=1,\ldots,m$}
                \For {$e=1,\ldots, \text{epochs}_{\calD}$}
                    \State Compute $\calL_{\calD}$ 
                    \State $\calW_j \gets \calW_j - \alpha_{\calD} \nabla\calL_{\calD}$
                \EndFor
            \EndFor
            \State Update $w_{i}, \forall{i}$ 
        \Until{convergence} 
	\end{algorithmic}
\end{algorithm}

\newpage
\section{Additional results}
\subsection{Effectiveness of resampling}

\begin{table}[h]
\caption{Results of ADC with/without the resampling process. Each AUC value represents the avg. AUC computed over 10 seeds. The higher performance between the resampling and no-resampling is indicated in bold.}
\label{tab:AUC-GENERALITY-ODD-NS-ADC}
\centering
\scriptsize
\begin{tabular}{c|c|cccccc}
\toprule
\multicolumn{2}{r|}{\begin{tabular}[c]{@{}r@{}}Dataset\\ Contamination ratio\end{tabular}} & \begin{tabular}[c]{@{}c@{}}Satellite\\ (31.6\%)\end{tabular} & \begin{tabular}[c]{@{}c@{}}Arrhythmia\\ (14.6\%)\end{tabular} & \begin{tabular}[c]{@{}c@{}}Cardio\\ (9.6\%)\end{tabular} & \begin{tabular}[c]{@{}c@{}}Shuttle\\ (7.2\%)\end{tabular} & \begin{tabular}[c]{@{}c@{}}Thyroid\\ (2.5\%)\end{tabular} & \begin{tabular}[c]{@{}c@{}}Satimage-2\\ (1.2\%)\end{tabular} \\ \midrule
\multirow{2}{*}{OCSVM-H}    & ADC-ns    &$\mathbf{71.9 \pm1.4}$     &$\mathbf{80.5 \pm0.8}$     &$\mathbf{92.6 \pm1.4}$     &$\mathbf{99.0 \pm0.1}$              &$96.5 \pm0.3$              &$\mathbf{99.6 \pm0.1}$ \\ 
                            & ADC       &$71.6 \pm1.2$              &$79.0 \pm0.8$              &$90.6 \pm1.8$              &$\mathbf{99.0 \pm0.0}$     &$\mathbf{96.6 \pm0.4}$     &$99.4 \pm0.2$          \\ \hline
                            
\multirow{2}{*}{AE}         & ADC-ns    &$\mathbf{63.2 \pm0.6}$     &$74.2 \pm0.1$              &$70.3 \pm0.5$              &$88.9 \pm14.6$             &$\mathbf{94.5 \pm0.8}$     &$89.4 \pm0.6$          \\ 
                            & ADC       &$\mathbf{63.2 \pm0.8}$              &$\mathbf{76.1 \pm0.4}$     &$\mathbf{71.1 \pm0.8}$     &$\mathbf{96.2 \pm3.0}$     &$93.0 \pm1.0$              &$\mathbf{90.2 \pm0.7}$ \\ \hline
                            
\multirow{2}{*}{Deep SVDD}  & ADC-ns    &$\mathbf{73.6 \pm4.8}$     &$75.0 \pm1.1$              &$60.5 \pm6.4$              &$73.2 \pm17.6$             &$77.7 \pm5.9$              &$84.4 \pm4.6$          \\ 
                            & ADC       &$72.5 \pm2.9$              &$\mathbf{77.7 \pm1.4}$     &$\mathbf{62.9 \pm4.7}$     &$\mathbf{74.9 \pm16.4}$    &$\mathbf{78.9 \pm3.2}$     &$\mathbf{87.5 \pm3.0}$ \\ \hline
                            
\multirow{2}{*}{RSRAE}      & ADC-ns    &$66.8 \pm0.7$              &$75.9 \pm0.4$              &$68.2 \pm1.1$              &$95.0 \pm2.3$              &$\mathbf{94.7 \pm0.7}$     &$90.1 \pm1.5$          \\ 
                            & ADC       &$\mathbf{67.1 \pm0.4}$     &$\mathbf{77.4 \pm0.3}$     &$\mathbf{69.3 \pm0.7}$     &$\mathbf{97.2 \pm1.7}$     &$94.4 \pm0.8$              &$\mathbf{90.2 \pm0.5}$ \\ \hline
                            
\multirow{2}{*}{DAGPR}      & ADC-ns    &$\mathbf{65.9 \pm3.4}$     &$81.7 \pm1.7$              &$\mathbf{92.0 \pm1.9}$     &N/A                        &$63.9 \pm6.3$              &$88.4 \pm5.6$          \\ 
                            & ADC       &$65.0 \pm1.5$              &$\mathbf{82.1 \pm1.2}$     &$90.6 \pm2.1$              &N/A                        &$\mathbf{66.3 \pm3.9}$     &$\mathbf{90.2 \pm3.7}$ \\ \bottomrule
\end{tabular}
\end{table}

\begin{table}[h]
\caption{Results of IADC with/without the resampling process. Each AUC value represents the avg. AUC computed over 10 seeds. The higher performance between the resampling and no-resampling is indicated in bold.}
\label{tab:AUC-GENERALITY-ODD-NS-IADC}
\centering
\scriptsize
\begin{tabular}{c|c|cccccc}
\toprule
\multicolumn{2}{r|}{\begin{tabular}[c]{@{}r@{}}Dataset\\ Contamination ratio\end{tabular}} & \begin{tabular}[c]{@{}c@{}}Satellite\\ (31.6\%)\end{tabular} & \begin{tabular}[c]{@{}c@{}}Arrhythmia\\ (14.6\%)\end{tabular} & \begin{tabular}[c]{@{}c@{}}Cardio\\ (9.6\%)\end{tabular} & \begin{tabular}[c]{@{}c@{}}Shuttle\\ (7.2\%)\end{tabular} & \begin{tabular}[c]{@{}c@{}}Thyroid\\ (2.5\%)\end{tabular} & \begin{tabular}[c]{@{}c@{}}Satimage-2\\ (1.2\%)\end{tabular} \\ \midrule
\multirow{2}{*}{OCSVM-H}    & IADC-ns    &$72.5 \pm1.5$              &$80.0 \pm1.6$              &$91.4 \pm0.5$              &$\mathbf{99.1 \pm0.1}$     &$\mathbf{96.6 \pm0.3}$     &$\mathbf{99.7 \pm0.0}$ \\ 
                            & IADC       &$\mathbf{75.4 \pm1.3}$     &$\mathbf{80.1 \pm1.7}$     &$\mathbf{94.0 \pm1.5}$     &$98.5 \pm0.0$              &$95.4 \pm0.6$              &$99.2 \pm0.1$          \\ \hline
                            
\multirow{2}{*}{AE}         & IADC-ns    &$61.9 \pm1.0$              &$74.2 \pm0.2$              &$71.4 \pm1.2$              &$93.9 \pm7.5$              &$95.0 \pm0.6$              &$90.8 \pm1.3$          \\ 
                            & IADC       &$\mathbf{66.5 \pm0.4}$     &$\mathbf{78.9 \pm0.1}$     &$\mathbf{71.4 \pm0.4}$     &$\mathbf{97.3 \pm2.0}$     &$\mathbf{95.9 \pm0.1}$     &$\mathbf{96.3 \pm1.5}$ \\ \hline
                            
\multirow{2}{*}{Deep SVDD}  & IADC-ns    &$72.0 \pm5.2$              &$75.9 \pm1.2$              &$64.4 \pm1.7$              &$82.4 \pm3.9$              &$78.4 \pm4.0$              &$81.0 \pm4.1$          \\ 
                            & IADC       &$\mathbf{82.2 \pm0.6}$     &$\mathbf{83.9 \pm0.5}$     &$\mathbf{75.1 \pm1.8}$     &$\mathbf{91.6 \pm10.4}$    &$\mathbf{86.8 \pm1.7}$     &$\mathbf{98.0 \pm2.8}$ \\ \hline
                            
\multirow{2}{*}{RSRAE}      & IADC-ns    &$67.3 \pm0.7$              &$76.1 \pm0.3$              &$68.4 \pm0.6$              &$\mathbf{98.0 \pm0.9}$     &$95.1 \pm0.5$              &$88.6 \pm0.6$          \\ 
                            & IADC       &$\mathbf{69.1 \pm0.2}$     &$\mathbf{80.0 \pm0.5}$     &$\mathbf{70.4 \pm0.6}$     &$97.2 \pm1.7$              &$\mathbf{95.9 \pm0.2}$     &$\mathbf{95.7 \pm2.5}$ \\ \hline
                            
\multirow{2}{*}{DAGPR}      & IADC-ns    &$\mathbf{68.4\pm4.6}$      &$81.2 \pm1.8$              &$93.0 \pm1.2$              &N/A                        &$63.1 \pm1.6$              &$87.8 \pm5.7$          \\ 
                            & IADC       &$67.3 \pm0.8$              &$\mathbf{82.0 \pm1.4}$     &$\mathbf{93.1 \pm1.6}$     &N/A                        &$\mathbf{65.0 \pm8.2}$     &$\mathbf{92.3 \pm1.4}$ \\ \bottomrule
\end{tabular}
\end{table}

In this experiment, we examine the performance with/without resampling process to show its effectiveness. The results in the case of not using resampling are indicated as `-ns (\textbf{n}o re\textbf{s}ampling)' in \Tref{tab:AUC-GENERALITY-ODD-NS-ADC} and \Tref{tab:AUC-GENERALITY-ODD-NS-IADC}. 
The experiment setting in this section is the same as in the main paper except for the use of resampling.

\Tref{tab:AUC-GENERALITY-ODD-NS-ADC} shows the performance according to the use of resampling in ADC.
In most base anomaly detectors, the performance is higher when resampling is used. 
 However, in the case of OCSVM or DAGPR, there are a few cases where the performance is higher when the ensemble base models use all data without resampling. 
This is presumably because OCSVM and DAGPR are based on kernel computation, thereby exploiting a full data distribution can benefit the individual base model prediction.
\Tref{tab:AUC-GENERALITY-ODD-NS-IADC} shows the performance according to the use of resampling in IADC. As in the ADC case, most base anomaly detectors perform better when resampling is used.
It is notable that the performance improvement from ADC to IADC is greater when the resampling is used.
%

\newpage
\subsection{Anomaly detection on unseen image data}


\begin{table*}[h]
\caption{\textcolor{red}{Results on MNIST test (unseen) datasets. Each AUC value represents the avg. AUC computed over 5 seeds for each of 10 anomaly detection scenarios. In each anomaly detector, the best performance among \textit{Single}, \textit{ADC}, and \textit{IADC} is indicated in bold.}} 
\label{tab:AUC-GENERALITY-IMG-UNSEEN}
\centering
\scriptsize
\begin{tabular}{c|c|cccc}
\toprule
\multicolumn{2}{r|}{\begin{tabular}[c]{@{}r@{}}Dataset\\ Contamination ratio\end{tabular}} & \multicolumn{1}{c}{\begin{tabular}[c]{@{}c@{}}MNIST\\ (20\%)\end{tabular}}   & \multicolumn{1}{c}{\begin{tabular}[c]{@{}c@{}}MNIST\\ (10\%)\end{tabular}}   & \multicolumn{1}{c}{\begin{tabular}[c]{@{}c@{}}MNIST\\ (5\%)\end{tabular}}              & \multicolumn{1}{c}{\begin{tabular}[c]{@{}c@{}}MNIST\\ (0)\end{tabular}} \\ \midrule
\multirow{3}{*}{\begin{tabular}[c]{@{}c@{}}OCSVM\\ (Best)\end{tabular}}
                                & Single &$84.0 \pm0.2$             &$88.5 \pm0.3$              &$91.2 \pm0.3$              &$90.7 \pm0.0$              \\
                                & ADC    &$84.1 \pm0.2$             &$88.7 \pm0.1$              &$91.3 \pm0.1$              &$90.8 \pm0.1$              \\
                                & IADC   &$\mathbf{84.9 \pm0.6}$    &$\mathbf{89.0 \pm0.2}$     &$\mathbf{91.4 \pm0.1}$     &$\mathbf{91.1 \pm0.1}$     \\ \hline

\multirow{3}{*}{\begin{tabular}[c]{@{}c@{}}OCSVM\\ (Worst)\end{tabular}}
                                & Single &$80.7 \pm0.7$             &$84.8 \pm1.0$              &$87.5 \pm0.8$              &$89.2 \pm0.0$              \\
                                & ADC    &$80.8 \pm0.4$             &$85.2 \pm0.3$              &$87.7 \pm0.4$              &$89.2 \pm0.0$              \\
                                & IADC   &$\mathbf{81.4 \pm0.4}$    &$\mathbf{86.5 \pm0.2}$     &$\mathbf{89.6 \pm0.3}$     &$\mathbf{89.3 \pm0.0}$     \\ \hline

\multirow{3}{*}{OCSVM-H}        & Single &$82.8 \pm1.2$             &$88.1 \pm0.8$              &$90.5 \pm1.4$              &$95.0 \pm0.8$              \\
                                & ADC    &$85.0 \pm0.5$             &$89.8 \pm0.4$              &$91.7 \pm0.5$              &$95.6 \pm0.5$              \\
                                & IADC   &$\mathbf{87.1 \pm0.4}$    &$\mathbf{90.9 \pm0.6}$     &$\mathbf{92.3 \pm0.4}$     &$\mathbf{96.0 \pm0.5}$     \\ \hline

\multirow{3}{*}{AE}             & Single &$79.9 \pm0.7$             &$85.3 \pm0.9$              &$88.1 \pm0.7$              &$92.4 \pm0.8$              \\
                                & ADC    &$80.9 \pm0.3$             &$85.8 \pm0.5$              &$88.5 \pm0.3$              &$92.4 \pm0.4$              \\
                                & IADC   &$\mathbf{82.1 \pm0.2}$    &$\mathbf{86.4 \pm0.2}$     &$\mathbf{89.1 \pm0.4}$     &$\mathbf{92.5 \pm0.3}$     \\ \hline

\multirow{3}{*}{Deep SVDD}      & Single &$79.8 \pm1.0$             &$85.0 \pm1.1$              &$88.0 \pm1.1$              &$92.4 \pm0.7$              \\
                                & ADC    &$84.0 \pm0.5$             &$89.4 \pm0.3$              &$92.1 \pm0.4$              &$95.2 \pm0.3$              \\
                                & IADC   &$\mathbf{84.5 \pm0.4}$    &$\mathbf{89.7 \pm0.7}$     &$\mathbf{92.4 \pm0.4}$     &$\mathbf{95.5 \pm0.2}$     \\ \hline

\multirow{3}{*}{RSRAE}          & Single &$83.2 \pm0.8$             &$87.4 \pm0.8$              &$89.7 \pm0.8$              &$92.7 \pm0.7$              \\
                                & ADC    &$84.1 \pm0.4$             &$87.9 \pm0.4$              &$90.0 \pm0.4$              &$92.3 \pm0.5$              \\
                                & IADC   &$\mathbf{84.9 \pm0.4}$    &$\mathbf{88.6 \pm0.3}$     &$\mathbf{90.0 \pm0.3}$     &$\mathbf{92.4 \pm0.5}$     \\ \hline

\multirow{3}{*}{DAGPR}          & Single &$77.5 \pm3.7$             &$76.7 \pm4.6$              &$75.7 \pm3.1$              &$68.1 \pm1.8$              \\
                                & ADC    &$80.8 \pm1.6$             &$80.9 \pm2.5$              &$77.6 \pm1.9$              &$67.9 \pm1.5$              \\
                                & IADC   &$\mathbf{81.1 \pm2.3}$    &$\mathbf{81.0 \pm1.3}$     &$\mathbf{79.0 \pm2.6}$     &$\mathbf{73.3 \pm1.2}$     \\ \bottomrule
\end{tabular}
\end{table*}
%

\begin{table*}[h]
\caption{\textcolor{red}{Results on CIFAR10 test (unseen) datasets. Each AUC value represents the avg. AUC computed over 5 seeds for each of 10 anomaly detection scenarios. In each anomaly detector, the best performance among \textit{Single}, \textit{ADC}, and \textit{IADC} is indicated in bold. OCSVM with ADC/IADC on CIFAR10 is omitted due to its heavy computational requirement.}}  
\label{tab:AUC-GENERALITY-IMG-UNSEEN}
\centering
\scriptsize
\begin{tabular}{c|c|cccc}
\toprule
\multicolumn{2}{r|}{\begin{tabular}[c]{@{}r@{}}Dataset\\ Contamination ratio\end{tabular}} & \multicolumn{1}{c}{\begin{tabular}[c]{@{}c@{}}CIFAR10\\ (20\%)\end{tabular}}   & \multicolumn{1}{c}{\begin{tabular}[c]{@{}c@{}}CIFAR10\\ (10\%)\end{tabular}}   & \multicolumn{1}{c}{\begin{tabular}[c]{@{}c@{}}CIFAR10\\ (5\%)\end{tabular}}              & \multicolumn{1}{c}{\begin{tabular}[c]{@{}c@{}}CIFAR10\\ (0)\end{tabular}} \\ \midrule
\multirow{3}{*}{\begin{tabular}[c]{@{}c@{}}OCSVM\\ (Best)\end{tabular}}
                                & Single &                           &                           &                          &                           \\
                                & ADC    &                           &N/A                        &                          &                           \\
                                & IADC   &                           &                           &                          &                           \\ \hline

\multirow{3}{*}{\begin{tabular}[c]{@{}c@{}}OCSVM\\ (Worst)\end{tabular}}
                                & Single &                           &                           &                          &                           \\
                                & ADC    &                           &N/A                        &                          &                           \\
                                & IADC   &                           &                           &                          &                           \\ \hline

\multirow{3}{*}{OCSVM-H}        & Single &$57.7 \pm3.7$              &$58.7 \pm2.8$              &$58.2 \pm2.6$             &$53.0 \pm0.2$              \\
                                & ADC    &$60.3 \pm1.8$              &$60.9 \pm1.2$              &$\mathbf{61.4 \pm0.5}$    &$54.0 \pm0.7$              \\
                                & IADC   &$\mathbf{60.3 \pm0.5}$     &$\mathbf{61.2 \pm1.4}$     &$61.4 \pm0.8$             &$\mathbf{54.1 \pm0.6}$     \\ \hline

\multirow{3}{*}{AE}             & Single &$53.8 \pm0.1$              &$54.4 \pm0.1$              &$54.9 \pm0.2$             &$55.4 \pm0.1$              \\
                                & ADC    &$53.9 \pm0.1$              &$\mathbf{54.6 \pm0.0}$     &$55.0 \pm0.0$             &$55.6 \pm0.1$              \\
                                & IADC   &$\mathbf{54.0 \pm0.0}$     &$54.6 \pm0.1$              &$\mathbf{55.1 \pm0.1}$    &$\mathbf{55.8 \pm0.0}$     \\ \hline

\multirow{3}{*}{Deep SVDD}      & Single &$58.7 \pm0.5$              &$60.7 \pm0.4$              &$61.8 \pm0.4$             &$63.2 \pm0.5$              \\
                                & ADC    &$\mathbf{59.6 \pm0.2}$     &$61.7 \pm0.3$              &$62.5 \pm0.3$             &$64.4 \pm0.3$              \\
                                & IADC   &$\mathbf{59.6 \pm0.2}$     &$\mathbf{62.0 \pm0.3}$     &$\mathbf{62.8 \pm0.3}$    &$\mathbf{64.5 \pm0.4}$     \\ \hline

\multirow{3}{*}{RSRAE}          & Single &$54.3 \pm0.1$              &$55.3 \pm0.1$              &$55.9 \pm0.1$             &$56.6 \pm0.1$              \\
                                & ADC    &$54.4 \pm0.0$              &$55.2 \pm0.1$              &$55.9 \pm0.1$             &$56.7 \pm0.2$              \\
                                & IADC   &$\mathbf{54.5 \pm0.1}$     &$\mathbf{55.4 \pm0.1}$     &$\mathbf{56.0 \pm0.1}$    &$\mathbf{56.9 \pm0.1}$     \\ \hline

\multirow{3}{*}{DAGPR}          & Single &$56.4 \pm1.2$              &$56.0 \pm1.0$              &$56.1 \pm1.1$             &$55.5 \pm1.1$              \\
                                & ADC    &$57.0 \pm0.8$              &$56.9 \pm0.8$              &$56.7 \pm0.7$             &$56.5 \pm0.4$              \\
                                & IADC   &$\mathbf{58.2 \pm2.7}$     &$\mathbf{57.1 \pm0.8}$     &$\mathbf{56.9 \pm0.6}$    &$\mathbf{57.2 \pm1.5}$     \\ \bottomrule
\end{tabular}
\end{table*}

In the main paper, we experiment with anomaly detection scenarios with transductive settings, \ie training and test on the same data. Here, we show anomaly detection scenarios with inductive settings where the test samples are unseen during the training (\Tref{tab:AUC-GENERALITY-IMG-UNSEEN}). For the inductive setting experiment, we sample both normal and anomalous samples from the test split of the image dataset while the anomaly detector only uses samples from the train split of the image dataset for the training. 
The performance on unseen test data is comparable to the performance of the transductive scenarios in the main paper. This validates that our framework not only enables a robust detection under data contamination but also generalizes to unseen data.

\newpage

\section{Additional analysis}
\subsection{Sensitivity analysis of the outlier ratio $r$}

\begin{figure}[H]
    \centering
    \subfigure[Satellite]{\includegraphics[width=0.3\textwidth]{Figures/inlier_satellite.png}} 
    \subfigure[Arrhythmia]{\includegraphics[width=0.3\textwidth]{Figures/inlier_arrhythmia.png}} 
    \subfigure[Cardio]{\includegraphics[width=0.3\textwidth]{Figures/inlier_cardio.png}}
    \subfigure[Shuttle]{\includegraphics[width=0.3\textwidth]{Figures/inlier_shuttle.png}}
    \subfigure[Thyroid]{\includegraphics[width=0.3\textwidth]{Figures/inlier_thyroid.png}}
    \subfigure[Satimage-2]{\includegraphics[width=0.3\textwidth]{Figures/inlier_satimage.png}}    
    \caption{The performance of IADC with varying outlier ratio $r$.}
    \label{fig:outlier}  
\end{figure}

\begin{figure}[H]
    \centering
    \subfigure[Satellite]{\includegraphics[width=0.3\textwidth]{Figures/auc_dsvdd_satellite.png}} 
    \subfigure[Arrhythmia]{\includegraphics[width=0.3\textwidth]{Figures/auc_dsvdd_arrhythmia.png}} 
    \subfigure[Cardio]{\includegraphics[width=0.3\textwidth]{Figures/auc_dsvdd_cardio.png}}
    \subfigure[Shuttle]{\includegraphics[width=0.3\textwidth]{Figures/auc_dsvdd_shuttle.png}}
    \subfigure[Thyroid]{\includegraphics[width=0.3\textwidth]{Figures/auc_dsvdd_thyroid.png}}
    \subfigure[Satimage-2]{\includegraphics[width=0.3\textwidth]{Figures/auc_dsvdd_satimage.png}}    
    \caption{The performance of Deep SVDD applied to IADC during iterative learning. The solid line and the colored part represent the average AUC and the standard deviation of AUC computed over 10 seeds. The red dotted line represents the average of the AUC obtained when the outlier ratio is 0.5.}
    \label{fig:auc-dsvdd}
\end{figure}

\newpage
The outlier ratio $r$ determines when to stop the iterative training. The performance along with different outlier ratios $r$ for the convergence criterion is plotted in \Fref{fig:outlier}. The ideal outlier ratio would be the ratio of anomalous samples in the contaminated dataset. However, the ratio $r$ is not sensitive to the performance. This result is interesting because the outlier hyperparameter $\nu$ used in OC-SVM variants is much more sensitive. This is because the hyperparameter $\nu$ directly affects the model parameter tunning, while the outlier ratio $r$ used in our approach only determines the moment of convergence.

We visualize the AUC mean and standard of IADC with Deep SVDD over iterations with 10 trials (\Fref{fig:auc-dsvdd}). The black line and dots represent the mean AUC at each iteration, and the colored region represents the standard deviation.
The dotted red line indicates the performance of IADC when the convergence criterion with outlier ratio $50\%$ is applied. 
As shown in \Fref{fig:auc-dsvdd}, our convergence criterion captures the moment when IADC performance saturates. Without the proper convergence metric, it is hard to decide when to stop the iterative process. Moreover, the proposed convergence criterion is particularly useful when the model's performance fluctuates as in the `Shuttle' scenario.

\subsection{Variance analysis of anomaly detection performance}

\begin{table}[H]
\caption{Average standard deviation of anomaly score for each data sample from multiple trials (10 seeds) in \textit{Single}, \textit{ADC}, and \textit{IADC}. The smallest average standard deviation among \textit{Single}, \textit{ADC}, and \textit{IADC} is indicated in bold. \textcolor{red}{DAGPR on Shuttle is ommitted due to its heavy computational requirement.}}
\label{tab:AUC-GENERALITY-ODD-STD}
\centering
\footnotesize
\begin{tabular}{c|c|cccccc}
\toprule
\multicolumn{2}{r|}{\begin{tabular}[c]{@{}r@{}}Dataset\\ Contamination ratio\end{tabular}} & \begin{tabular}[c]{@{}c@{}}Satellite\\ (31.6\%)\end{tabular} & \begin{tabular}[c]{@{}c@{}}Arrhythmia\\ (14.6\%)\end{tabular} & \begin{tabular}[c]{@{}c@{}}Cardio\\ (9.6\%)\end{tabular} & \begin{tabular}[c]{@{}c@{}}Shuttle\\ (7.2\%)\end{tabular} & \begin{tabular}[c]{@{}c@{}}Thyroid\\ (2.5\%)\end{tabular} & \begin{tabular}[c]{@{}c@{}}Satimage-2\\ (1.2\%)\end{tabular} \\ \midrule
\multirow{3}{*}{OCSVM}      & Single    & 0.0247            & 0.0433            & 0.0398            & 0.0263            & 0.0318            & 0.0351            \\
                            & ADC       & 0.0131            & \textbf{0.0216}   & 0.0205            & 0.0142            & 0.0160            & 0.0152            \\
                            & IADC      & \textbf{0.0110}   & 0.0266            & \textbf{0.0194}   & \textbf{0.0112}   & \textbf{0.0157}   & \textbf{0.0145}   \\ \hline
                            
\multirow{3}{*}{AE}         & Single    & 0.0176            & 0.0079            & 0.0273            & 0.0755            & 0.0524            & 0.0168            \\
                            & ADC       & 0.0071            & 0.0053            & 0.0138            & 0.0397            & 0.0235            & 0.0069            \\
                            & IADC      & \textbf{0.0058}   & \textbf{0.0037}   & \textbf{0.0083}   & \textbf{0.0217}   & \textbf{0.0053}   & \textbf{0.0056}   \\ \hline
                            
\multirow{3}{*}{Deep SVDD}  & Single    & 0.0841            & 0.0527            & 0.0942            & 0.0981            & 0.0952            & 0.0948            \\
                            & ADC       & 0.0395            & 0.0276            & 0.0497            & 0.0525            & 0.0439            & 0.0477            \\
                            & IADC      & \textbf{0.0073}   & \textbf{0.0033}   & \textbf{0.0222}   & \textbf{0.0451}   & \textbf{0.0052}   & \textbf{0.0256}   \\ \hline
                            
\multirow{3}{*}{RSRAE}      & Single    & 0.0124            & 0.0099            & 0.0257            & 0.0592            & 0.0455            & 0.0127            \\
                            & ADC       & 0.0057            & 0.0064            & 0.0132            & 0.0275            & 0.0230            & 0.0053            \\
                            & IADC      & \textbf{0.0035}   & \textbf{0.0056}   & \textbf{0.0101}   & \textbf{0.0141}   & \textbf{0.0068}   & \textbf{0.0041}   \\ \hline
                            
\multirow{3}{*}{DAGPR}      & Single    & 0.0647            & 0.0722            & 0.0865            &                   & 0.0541            & 0.0634            \\
                            & ADC       & 0.0370            & 0.0385            & \textbf{0.0304}   &N/A                & 0.0239            & \textbf{0.0265}   \\
                            & IADC      & \textbf{0.0250}   & \textbf{0.0327}   & 0.0316            &                   & \textbf{0.0198}   & 0.0286            \\ \bottomrule
\end{tabular}
\end{table}

The contaminated anomaly detection scenario implies a noisy data distribution. As a result, the trained model may learn a less reliable representation. In such a case, anomaly scores of individual samples are likely to have high variances over multiple trials. We show that the anomaly score standard deviation of a single anomaly detection model is significantly larger than that of ADC or IADC (\Tref{tab:AUC-GENERALITY-ODD-STD} and \Fref{fig:std-dsvdd}). We measure the variance of anomaly scores as follows.

\begin{equation}
    \label{eqn:anomaly-std}
    Var(s_i) = \frac{1}{b}\sum_{k=1}^b{(\sigma(s_i^k)-\bar{s_i})^2}, \quad \text{where} \quad \bar{s_i}=\frac{1}{b}\sum_{k=1}^b{\sigma(s_i^k)}
\end{equation}

where $k$ is an index of trials, $b$ is the number of total trials, $s_i^k$ is an anomaly score of data $x_i$ at $k^{th}$ trial, and $\sigma$ is a sigmoid function. We apply a sigmoid function on anomaly scores to map values into the range $[0,1]$ only for variance analysis because the scales of anomaly scores across the models are different.
In \Tref{tab:AUC-GENERALITY-ODD-STD}, the numbers represent an average of standard deviations of anomaly scores. The standard deviation of ADC and IADC is significantly lower than the single model verifying that the proposed framework outputs a consistent result when noisy data is given. For better visualization of the results, we draw boxplots of the standard deviations of single, ADC and IADC with Deep SVDD in \Fref{fig:std-dsvdd}. 

\begin{figure}[H]
    \centering
    \subfigure[Satellite]{\includegraphics[width=0.3\textwidth]{Figures/std_dsvdd_satellite.png}} 
    \subfigure[Arrhythmia]{\includegraphics[width=0.3\textwidth]{Figures/std_dsvdd_arrhythmia.png}} 
    \subfigure[Cardio]{\includegraphics[width=0.3\textwidth]{Figures/std_dsvdd_cardio.png}}
    \subfigure[Shuttle]{\includegraphics[width=0.3\textwidth]{Figures/std_dsvdd_shuttle.png}}
    \subfigure[Thyroid]{\includegraphics[width=0.3\textwidth]{Figures/std_dsvdd_thyroid.png}}
    \subfigure[Satimage-2]{\includegraphics[width=0.3\textwidth]{Figures/std_dsvdd_satimage.png}}    
    \caption{Boxplots of the standard deviations of anomaly scores for each data sample from multiple trials (10 seeds) in \textit{Single}, \textit{ADC}, \textit{IADC} with Deep SVDD.}
    \label{fig:std-dsvdd}
\end{figure}

\subsection{Visualization of importance weights and feature distributions}
In this subsection, we show additional visualizations for changes of importance weights and feature distributions using Deep SVDD~\cite{ruff2018deep} (\Fref{fig:visualize_weights}).
The left graphs visualize the changes of importance weights up to 30 iterations. The average and median values of importance weights of normal and anomalous samples show a clear gap.
The right feature distribution is visualized by t-SNE~\cite{van2008visualizing} showing the effectiveness of our framework compared to a single anomaly detector. 
The feature space of our approach shows a more distinctive separation between normal and anomalous samples. For the t-SNE visualization, we use the IADC model trained up to convergence using the stopping rule introduced in the main paper.

\begin{figure}[h!]
  \centering
  \includegraphics[width=0.9\linewidth]{Figures/supp-satellite.png} 
  \includegraphics[width=0.9\linewidth]{Figures/supp-cardio.png} 
  \includegraphics[width=0.9\linewidth]{Figures/supp-thyroid.png} 
  \includegraphics[width=0.9\linewidth]{Figures/supp-satimage.png} 
  \caption{\textcolor{red}{
  \textbf{(Left)} Changes in importance weights in IADC with Deep SVDD. From top to bottom, the visualized datasets are Satellite, Cardio, Thyroid, and Satimage-2. Here, the median and mean of the importance weights of normal and abnormal data of each dataset over iterations are shown. 
  The colored part represents the standard deviation of the importance weights of each data group. 
  \textbf{(Middle \& Right)} (Upper) The distribution of features and (Lower) the histogram of the anomaly score obtained from (Middle) single Deep SVDD and (Right) IADC. t-SNE is used to visualize the features in two-dimensional space, where the dashed circles represent regions containing 25\%, 50\%, and 75\% of the dataset from the center of the hypersphere, which is noted as the black 'x' mark. 
  The thin dashed line over each histogram is a probability density function fitted by the Gaussian distribution of anomaly score.
  The thick dashed line is the center of each Gaussian distribution.}}
  \label{fig:visualize_weights}
\end{figure}

\clearpage
\bibliography{mybib}